\documentclass[12pt]{article}
\usepackage{amsmath, setspace}
\usepackage{graphicx}
\usepackage{lineno}

 \makeatletter       
 \renewcommand{\@biblabel}[1]{#1.}
 \makeatother
\begin{document}
\noindent
{\LARGE\textbf{Angle of repose of Martian wet sand using discrete element method: Implication for the seasonal cycle of recurring slope lineae(RSL) by relative humidity}}
\\
\\
{\large Daigo Shoji$^{1}$(shoji@elsi.jp), Shoko Imamura$^{1}$, Maya Nakamura$^{1}$, Rina Noguchi$^{2}$}\\
\\
1. Earth-Life Science Institute, Tokyo Institute of Technology, 2-12-1 Ookayama, Meguro-ku, Tokyo\\
2. Institute of Space and Astronautical Science, Japan Aerospace Exploration Agency, 3-1-1 Yoshinodai, Chuo-ku, Sagamihara, Kanagawa\\
\newpage
\section*{Abstract}
Seasonal cycle of Martian relative humidity shows that, in the southern and the northern RSL sites, humidity decreases when RSL increment while humidity is high on the fading season. We suggest that the recurrence of RSL is caused by the different angle of response between wet and dry grains. 2D DEM simulation indicates that, when the ratio of a liquid bridge volume to a particle volume is the order of 10$^{-7}$, the repose angle of Martian sand can increase to $\sim$40 degrees due to cohesion, which is near the slope around RSL starting area. Thus, if this amount of liquid bridge is generated by deliquescence or capillary condensation of vapor and if the water evaporates following the humidity and temperature cycles, Martian sand can induce avalanche on RSL slope. When grains are wet again, sand and dust deposit on RSL, which may be the mechanism of RSL disappearance and recharge of grains for the next season. In our model, because liquid induces RSL indirectly (RSL itself is the granular flow), the amount of liquid can be drastically reduced compared with the wet mechanism. In this case, as well as temperature, the seasonal cycle of relative humidity plays an important role in the recurrence of RSL. 

\section{Introduction}
On Mars, it has been observed that dark lineae lengthen and fade cyclically every Martian year (e.g., McEwen et al., 2011, 2014; Stillman et al., 2014, 2016, 2017), which is called recurring slope lineae (RSL). Regardless of many researches, the recurrence mechanism of RSL is controversial. One effective hypothesis is that RSL is caused by liquid water (wet mechanism). Analyses of images revealed that RSL appear with increasing temperature (e.g., Stillman et al., 2014, 2016, 2017). Although pure water is not stable for Martian environment, several salts can reduce the melting point and the evaporation rate, which stabilizes the liquid (e.g., Chevrier et al., 2012; Gough et al., 2019). The wet mechanism can explain seasonal cycle of RSL smoothly. Low albedo of lineae is also consistent with the appearance of liquid (ground becomes dark by wetting). However, direct evidence of water is not found in RSL so far although the existence of hydrated salts is indicated (Ojha et al., 2015). In addition, the wet mechanism requires relatively a large amount of liquid to generate RSL. Because water supply only from vapor is not sufficient (Chevrier et al., 2012; Dundas et al., 2017), water budget such as briny aquifer is required (Stillman et al., 2016). Such a large aquifer is not found so far under shallow depth (Wilson et al., 2018), and further observations are required.

The other hypothesis suggests that RSL are caused by flow or removal of sand or dust (dry mechanism). It has been observed that many RSL stop at the slope near the angle of repose of dry sand ($\sim$30 degrees) regardless of their length (Schmidt et al., 2017; Dundas et al., 2017), which supports the granular flow. The dry mechanism does not require water budget. However, this mechanism is challenged to the seasonal recurrence of RSL. Although several mechanisms are suggested for the generation of RSL such as Knudsen pump (Schmidt et al., 2017) or dust removal by wind (Vincendon et al., 2019), it is still uncertain how RSL appear every year. Fading mechanism is also an enigma from the aspect of dry mechanism.

Recently global distributions of Martian relative humidity were evaluated by P\'al et al. (2019). Considering their results, at the RSL sites in the southern mid-latitude, relative humidity decreases from $L_s$=180$^\circ$ to 270$^\circ$, and increases from $L_s$=270$^\circ$ to 360$^\circ$. In the case of RSL sites in the northern hemisphere (Chryse and Acidalia Planitiae), humidity decreases between $L_s$=270$^\circ$ and 360$^\circ$, and increases from $L_s$=180$^\circ$ to 270$^\circ$. The evaluation by Davila et al. (2010) also indicates that, in the southern mid-latitude, high humidity season is between $L_s$=0$^\circ$ and 200$^\circ$, which is anti-correlated with temperature (temperature is high when humidity is low). In both hemispheres, relative humidity decreases and increases when RSL begin to lengthen and fade, respectively (Stillman et al., 2014, 2016; Vincendon et al., 2019). Around the equatorial region, although the seasonal cycle is different between each RSL, several sites such as Garni crater have RSL that increment between $L_s$=90$^\circ$ and 180$^\circ$ (Stillman et al., 2017), which is the season when relative humidity decreases (P\'al et al.,2019; Martin-Torres et al., 2015). This link between RSL recurrence and humidity cycles may show that humidity as well as temperature drives RSL at least in the northern and the southern mid-latitude. The effect of water vapor through deliquescence has been suggested (e.g., Heinz et al., 2016). However, the global maps show that humidity decreases when RSL increment (P\'al et al.,2019), which is not consistent that vapor itself is the material of RSL.

On the Earth, we can see sand avalanche  due to drying at Tottori Sand Dune in Japan. When it rains, wet sand is blown and piles on the top of dune's slope. After the raining, sand dries and flows downward (Bitou and Kodama, 2011). This sand flow makes lineae called "Saren" (sand curtain). The timing of Saren is linked to the season (i.e., climate) because both rainy (wetting) and sunny (drying) days are required (Bitou and Kodama, 2011). 

In this work, as mentioned in the next section, we suggest that the wetting and the drying of grains (sand or dust) by the seasonal humidity and temperature changes may be the cause of the recurrence of RSL. Due to a cohesion, wet particles have a lager angle of response (critical and repose angles) compared to dry particles (e.g., Mitarai et al., 2006). Thus wet grains can flow to downslope as they dry. When humidity increases again, particles can be wet and deposit on RSL. In this scenario, liquid works to erase RSL rather than the generation of lineae, which is consistent with the humidity cycle. Using discrete elemental method (DEM), we calculate the angle of repose of wet particles with different liquid volume. Here we show preliminary results of DEM simulations and future works.

\section{Angle of repose and RSL cycle}
When a slope of granular pile increases to the critical angle $\theta_c$, grains begin to flow until the slope becomes $\theta_r$, which is called the angle of repose (Liu et al., 2005). Several experiments revealed that $\theta_r$ usually becomes smaller than $\theta_c$ by a few degrees (e.g., Liu et al., 2005). In the case of dry sand, $\theta_r$ usually becomes $\sim$30 degrees (e.g., Liu et al., 2005). Many RSL stop at $\sim$30 degrees of slope (Schmidt et al., 2017; Dundas et al., 2017). If this angle shows $\theta_r$ of Martian dry sand, the Martian repose angle is almost the same as that of terrestrial grains. 

Once particles are wet, both $\theta_c$ and $\theta_r$ increase due to the cohesion by a capillary force (e.g., Hornbaker et al., 1997; Nowak et al., 2005; Samadani and Kudrolli, 2001). If we consider a toroidal liquid bridge between grains, cohesion force $f_c$ is given as 

\begin{equation}
f_c\sim2\pi r_N\gamma +\pi r_N^2\Delta P,
\label{cohesion1}
\end{equation}
where $r_{N}$, $\gamma$ and $\Delta P$ are radius of liquid bridge's neck, surface tension and pressure difference between water and air, respectively (Mitarai et al., 2006). Several experiments show that $\theta_r$ increases with the volume of liquid bridge and then becomes stable due to the saturation of liquid (Samadani and Kudrolli, 2001). In order for the repose angle to rise by $\sim$10 degrees, only a few tens of nano meters of liquid coating is sufficient (Hornbaker et al., 1997; Albert et al., 1997; Tegzes et al., 1999), which is approximately 10$^{-4}$ of the grain volume. On Mars, although liquid pure water is unstable, several salts have been suggested to induce deliquescence by reducing the melting point and the evaporation rate of brine (e.g., Chevrier et al., 2012; Gough et al., 2014, 2019). In addition, even though a sufficient amount of salt is not contained in the soil, pure water can make liquid bridge (capillary condensation) once particles come into contact at more than $\sim$40\% of relative humidity, which results in high repose angle  (Culligan and Christenson, 2014; G\'omez-Arriaran et al., 2015). If the curvature of liquid bridge's surface (meniscus) is nano scale, water between particles can be meta-stable liquid phase even under minus pressure and lower temperature than the melting point (Yang et al., 2008; Nosonovsky and Bhushan, 2008). Particles have surface roughness with $\sim$1 $\mu$m of length scale (Hornbaker et al., 1997; Mason et al., 1999). Water around this size of tip can remain in liquid phase as the meta-stable state (Yang et al., 2008). 

The evaporation rate of liquid depends on the diurnal cycle of temperature and humidity as well as the seasonal change. In daytime, due to the high temperature and the low humidity (Martin-Torres et al., 2015), generated liquid may evaporate completely. However, the thickness of sand flow on dunes is a few centimeters (Dundas et al., 2017). Below a few centimeters depth, the evaporation rate is largely reduced due to the low temperature, and water is more stable compared with surface (Chevrier et al., 2012; Martin-Torres et al., 2015; Primm et al., 2019). Although the liquid freezes in the night and on winter season, frozen liquid can maintain the cohesion. Thus, when the seasonal relative humidity is high, $\sim$40 degrees of repose angle may be possible due to the liquid bridge. 

\begin{figure}[htbp]
\centering
\includegraphics[width=19cm] {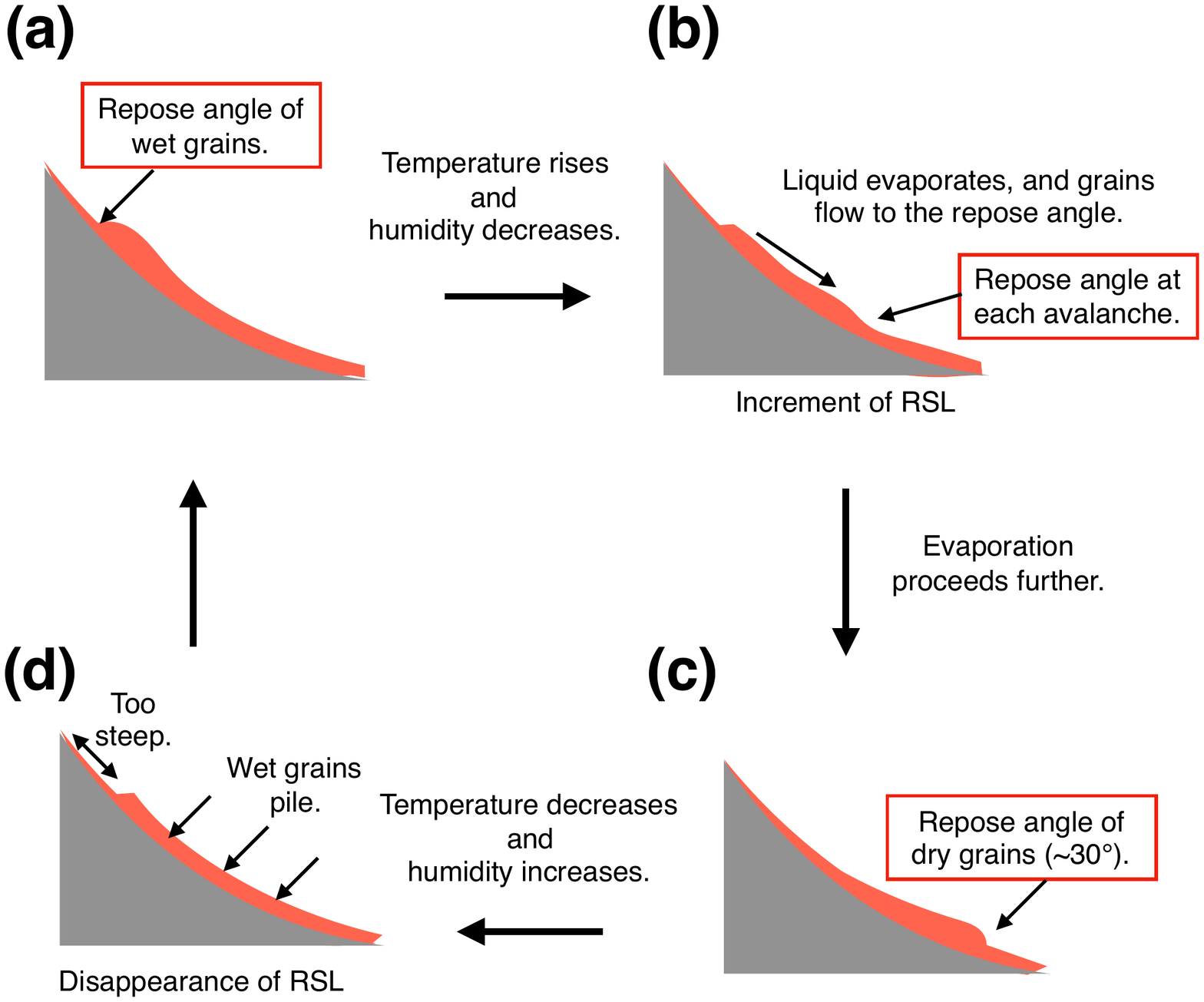}
\caption{Schematic view of RSL recurrence. Gray and red areas show the side view of bedrock and sand, respectively. (a): When humidity is high, transported wet grains pile on the slope up to the repose angle. Due to the flow from upslope, the thickness of sand is larger around the repose angle. (b): As relative humidity decreases and temperature increases, due to the decrease of critical angle, grains around thick layer begin to flow to the repose angle, which induces RSL. If evaporation is not complete, RSL lengthen gradually depending on both the critical and the repose angles at each time. Overprinting due to the flow of new exposed surface on upslope also occurs. (c): Finally the avalanche stops at the repose angle of dry sand ($\sim$30 degrees of slope). (d): When humidity increases and temperature is reduced again, transported particles are wet and deposit on slope up to the source point, which erases RSL and recharges grains for next season.}
\label{view}
\end{figure}

Fig. \ref{view} shows the schematic view on the recurrence mechanism of RSL suggested in this work. Although the deviation is large, it has been indicated that RSL start from approximately or less than 40 degrees (McEwen et al., 2011; Schmidt et al., 2017; Schaefer et al., 2019; Tebolt et al., 2019). We suppose that this angle is the angle of repose of Martian wet sand. At high humidity season, wet grains transported from surrounding area pile on the area whose slope is less than $\theta_r$ without flowing (Fig. \ref{view} a). On the upper area with slope $>\theta_r$, grains cannot pile even though it is wet and they flow to the area with slope around $\theta_r$. Thus, the thickness of granular layer is large around $\theta_r$ of slope (granular clump is generated), which triggers the sand avalanche. Many RSL start from outcrop located on upslope. Morphologic roughness of the bedrock can be the area of granular clump. 

When the seasonal humidity decreases and the seasonal temperature increases (from spring to summer season), liquid in shallow depth begins to evaporate gradually. Due to the steep slope and the large thickness, avalanche is induced from upslope (Fig. \ref{view} b). 
For example, when $\theta_c$ decreases by evaporation to $\sim$40 degrees (i.e., $\theta_r$ at the initial state), grains can flow to $\sim$35-37 degrees (i.e., $\theta_r$ at this time). Grains on the slope smaller than $\theta_r$ ($<$35 degrees) do not move at this time. Due to the avalanche from the upslope, the thickness around $\theta_r$ increases (Fig. \ref{view} b). As seasonal temperature reaches maximum, evaporation proceeds and $\theta_c$ decreases further, which induces the next avalanche to the downslope. In addition to the downward avalanche, on upslope, buried grains are exposed to the surface by removing the former surface, which induces the further flow from upslope. Thus, overprinting on the generated RSL and supply of grains also occur at this stage. Finally, when grains are sufficiently dried, grains flow to $\theta_r$ for dry sand, which is $\sim$30 degrees (Fig. \ref{view} c). This gradual flowing process of sand can be regarded as the increment of RSL. 
It is observed that RSL incrementation has two or three pulses in the southern mid-latitude, in which RSL lengthen largely (Stillman and Grimm, 2018). This discontinuous incrementation may be caused by the difference between $\theta_c$ and $\theta_r$. Once grains flow to $\theta_r$, they do not move until $\theta_c$ is reduced to that slope. Thus avalanche proceeds discontinuously regardless of continuous evaporation.
When the seasonal temperature is reduced and the seasonal humidity increases, liquid can be stable again at the shallow depth. Thus particles (especially small grains such as dust) can pile on the RSL up to its source point, which erases RSL and recharges grains for the RSL in next season (Fig. \ref{view} d). 

In the following section, in order to evaluate how much liquid is required to increase the repose angle at Martian gravity, we calculate the angle of repose of wet grains using DEM simulations.


\section{Method}
\subsection{Discrete element method}
For the 2D DEM simulations, we used MercuryDPM (http://www.mercurydpm.org), which is the calculation software for DEM (Thornton et al., 2012; Weinhart et al., 2012). DEM models the dynamics between particles using degrees and dash-pot. If $i$th particle interacts with $j$th particle, equations of motion for translation and rotation can be given (Liu et al. 2011) as 

\begin{equation}
m_i\ddot{\vec{r}}_i=m_i\vec{g}+\sum_j(\vec{f}_n+\vec{f}_t+\vec{f}_c)
\end{equation}
\begin{equation}
I_i\dot{\vec{\omega}}_i=\sum_j(\vec{R}_i\times \vec{f}_t-R_i |\vec{f}_r| \hat{\omega}_i),
\end{equation}
where $m_i$, $\vec{r}_i$, $I_i$, $\vec{\omega}_i$, $\vec{g}$, $R_i$ are mass, position, moment of inertia, angular velocity, gravitational acceralation and radius of $i$th particle, respectively. $\vec{R}_i$ and $\hat{\omega}$ are the vector from the center of $i$th particle to the contact point and the unit vector to the direction of $\vec{\omega}$. $\vec{f}_n$, $\vec{f}_t$, $\vec{f}_r$ and $\vec{f}_c$ are normal force, tangential force, rolling resistance and cohesion force, respectively. Considering a Voigt model, in which degrees and dash-pot are connected in parallel (e.g. Nakashima et al., 2011), the magnitude of the normal force $f_n$ is given (e.g., Luding et al., 2017) by

\begin{equation}
f_n=k\delta_n+\eta v_n,
\end{equation}
where $k$, $\eta$, $\delta_n$ and $v_n$ are spring constant, damping coeffient, displacement and relative velocity to the normal direction of $i$th particle, respectively. In the case of the tangential force and the rolling resistance, frictional forces are added to the particle. Thus the magnitudes of the tangential and the rolling forces $f_t$ and $f_r$ are given  (Luding et al., 2017) by 

\begin{equation}
f_t=\mathrm{min}(\mu f_n, k_t\delta_t+\eta_t v_t)
\end{equation}

\begin{equation}
f_r=\mathrm{min}(\mu_r f_n, k_t\delta_t+\eta_t v_t),
\end{equation}
where $k_t$, $\eta_t$, $\delta_t$ and $v_t$ are spring constant, damping coefficient, displacement and relative velocity to the tangential direction of $i$th particle, respectively. $\mu$ and $\mu_r$ are the coefficients of sliding and rolling frictions, respectively. 

\subsection{Effect of cohesion}
Although the equation of the adhesion force can be given by Eq. (\ref{cohesion1}) considering the simple geometry of liquid bridge, the behavior of liquid between grains depends on many effects such as surface roughness and structure of liquid bridge. Thus, in this work, we take into account the equation derived from experimental results by Willett et al. (2000) as 

\begin{equation}
f_c=\frac{2\pi\gamma R \cos\phi}{1+2.1(S\sqrt{R/V})+10.0(S\sqrt{R/V})^2},
\label{cohesion2}
\end{equation}
where $\phi$, $S$ and $V$ are contact angle, half separation between two particles and volume of one liquid bridge, respectively. This equation is valid as long as $V/R_{ij}^3$ is less than 10$^{-3}$, where $R_{ij}$ is the harmonic mean radius of $i$th and $j$th particles (Willett et al, 2000). As shown below, we consider the liquid volume range around $V/R_{ij}^3$$\sim$10$^{-7}$, which is sufficiently small to use Eq. (\ref{cohesion2}). Following Eq. (\ref{cohesion2}), as particles come closer ($S$ decreases), the cohesion force can increase to $2\pi\gamma R \cos\phi$. However, actual particles have surface roughness with $\sim$1 $\mu$m of length scale (Hornbaker et al., 1997; Mason et al., 1999), which limits the increase of cohesion force. In our simulations, we assumed that $f_c$ does not exceed $f_c^{max}$, which was defined as $f_c$ with $S=0.5$ $\mu$m (separation 2$S$ is 1 $\mu$m). Thus the cohesion force was calculated as $f_c^{max}$ when separation 2$S$ is less than 1 $\mu$m.

The liquid bridge ruptures when particles are separated sufficiently. The critical half separation where cohesion force can work $S_c$ is evaluated (Lian et al., 1993; Willett et al., 2000) as 

\begin{equation}
 2S_c\sim\left(1+\frac{\phi}{2}\right)V^{1/3}.
 \label{critical}
 \end{equation}
This critical separation determines whether $f_c$ is induced between each grain. 

\subsection{Calculation of angle of repose}
\begin{figure}[htbp]
\centering
\includegraphics[width=17cm] {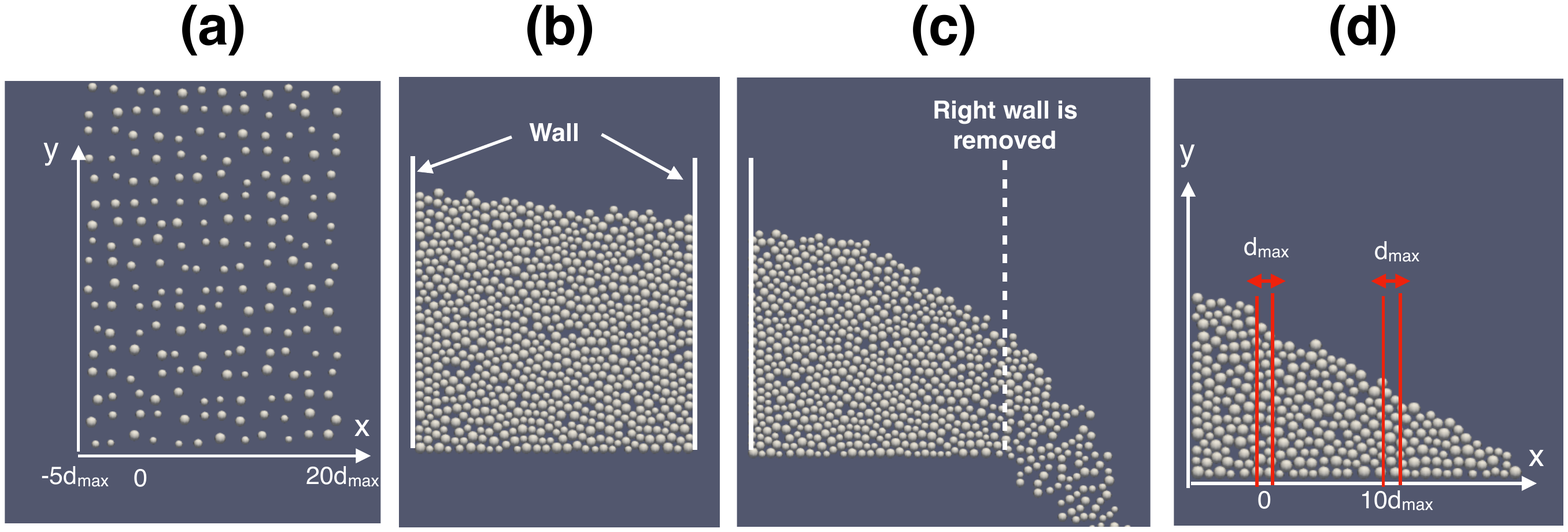}
\caption{Calculation process of 2D DEM simulation. (a): In 2$d_{max}$$\times$2$d_{max}$ grids, particles are generated. (b): Every particle is piled as a free fall within the two walls. (c): After every particle stops, right wall is removed, which causes the flow of particles. (d): After the flow of particles, the two slopes are calculated from the positions of the highest particles within $d_{max}$ from $x=0$ and $x$=10$d_{max}$. The angle of repose is evaluated as the mean slope of the calculated two slopes.}
\label{method}
\end{figure}

Firstly particles are generated in the area bounded by two walls (Fig. \ref{method} a). The width of the area $L$ was set at $L=25d_{max}$ between $x=-5d_{max}$ and $x=20d_{max}$, where $d_{max}$ was the maximum diameter of particles. The extra width of the left side (5$d_{max}$) was set to remove the effect of the left wall. Each particle was set within the 2$d_{max}\times$2$d_{max}$ grid (Fig. \ref{method} a). Position of particle's center in each grid was determined with random number. Then every particle was piled in the domain as a free fall (Fig. \ref{method} b). After every particle stops, the right wall was removed, which results in avalanche of the grains (Fig. \ref{method} c). 

Angle of repose $\theta_r$ can be calculated as 

\begin{equation}
\theta_r(x)=\tan^{-1}\left(\frac{h(x)}{L-x}\right),
\end{equation}
where $h(x)$ is the height of slope at $x$. Because the surface of grains is not flat, $\theta_r$ depends on $x$. Here we calculated the angel of repose as the mean angle of the slopes at $x\approx$ 0 and $x\approx$ 10$d_{max}$. The values of $h(x)$ and $x$ were determined from the position of the highest particles located within $x\pm$$d_{max}$/2, respectively (the particles whose centers were within $d_{max}$ from $x$ = 0 and 10$d_{max}$) (Fig. \ref{method} d). 

\begin{figure}[htbp]
\centering
\includegraphics[width=12cm] {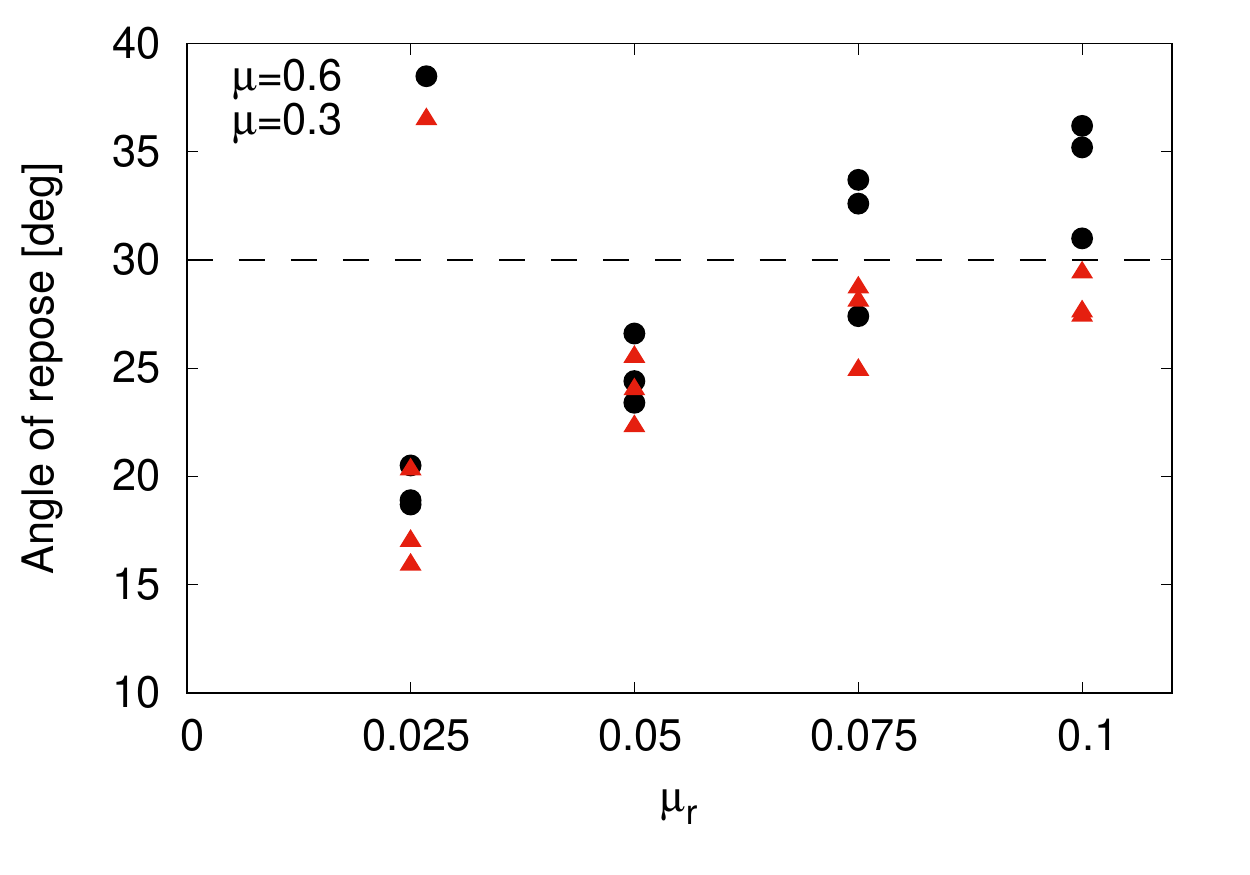}
\caption{Angle of repose of dry particles with different sliding and rolling coefficients. Diameter of particles are between 200 $\mu$m and 300 $\mu$m, and $g$=3.7 ms$^{-2}$. The horizontal line shows the angle of repose (30 degrees) for Martian dry sand (slope where RSL stop).}
\label{fig1}
\end{figure}

Values of each parameter are given in Table \ref{tab1}. Diameter of each grain was determined by random number within 50 $\mu$m from each mean diameter, and we tried the calculations with different mean diameters of particles. The most important parameters that affect the angle of repose are the sliding and rolling friction coefficients $\mu$ and $\mu_r$. On Mars, many RSL stop at $\sim$30 degrees (Schmidt et al., 2017; Dundas et al., 2017). Thus, using dry particles, we constrained the values of $\mu$ and $\mu_r$ considering that $\sim$30 degrees is the angle of repose of dry sand on Mars. Fig. \ref{fig1} shows the angle of repose at Martian gravity with different $\mu$ and $\mu_r$ assuming 200-300 $\mu$m of grain diameter, which is consistent with the size of Martian sand (Mass\'e et al., 2016). Calculations were performed three times with each set of the coefficients. The repose angle increases with increasing both coefficients because the magnitude of frictional force rises, which prevents particles from flowing. Although $\mu$ and $\mu_r$ cannot be determined uniquely only from one value ($\sim$30 degrees of $\theta_r$), considering that the terrestrial sand has $\mu$$\sim$0.6 (Nakashima et al., 2011), $\sim$0.075 of $\mu_r$ is consistent with the angle of repose at RSL sites. In the case of dry particle, we confirmed that the angle of repose little depends on mean diameter within the range of a few hundreds of micro meter. Thus we use $\mu=0.6$ and $\mu_r=0.075$ for the calculations of wet particles. 

\section{Result}
\begin{figure}[t]
\centering
\includegraphics[width=12cm] {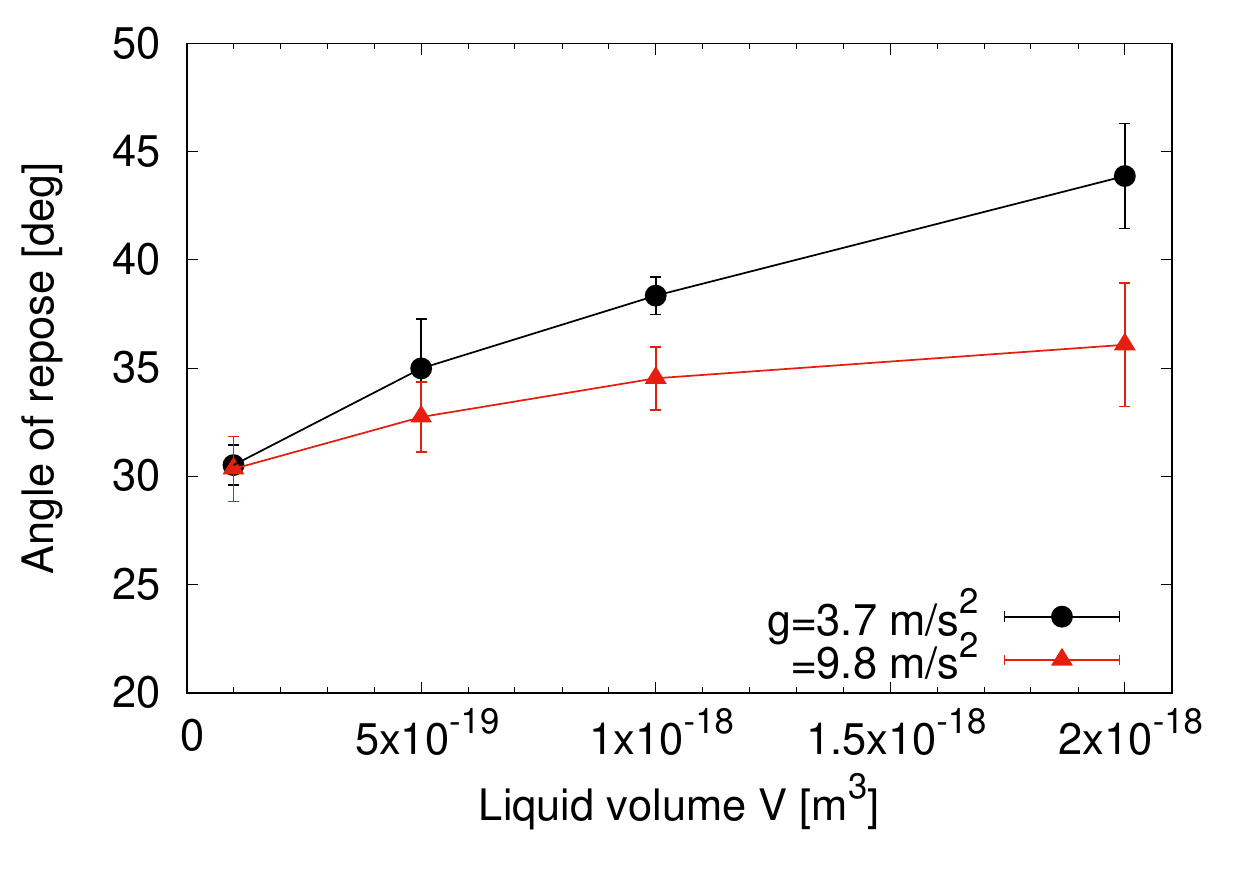}
\caption{Angle of repose of wet particles as a function of liquid bridge volume. Diameter of particles are between 200 $\mu$m and 300 $\mu$m. Circles show the repose angle on Mars ($g$=3.7 ms$^{-2}$) and triangles are the values on the Earth ($g$=9.8 ms$^{-2}$). Error bars are 1$\sigma$ of standard deviation of 10 calculations.}
\label{fig2}
\end{figure}

Fig. \ref{fig2} shows the angle of repose as a function of liquid bridge volume $V$ with the different gravities. The particle diameter is 200-300 $\mu$m (mean diameter is 250 $\mu$m). The repose angles were averaged by 10 calculations with the same parameter set. Error bars are 1$\sigma$ of standard deviation. At the Earth's gravity ($g$=9.8 ms$^{-2}$), the averaged angle of repose increases to 36 degrees at $V$=2$\times$10$^{-18}$ m$^{3}$, which is 2.4$\times$10$^{-7}$ of particle volume. In the case of experiments by Hornbaker et al. (1998) using polystyrene beads with 0.8 mm of mean diameter, the repose angle increases by $\sim$7-8 degrees when $V$$\sim$3$\times$10$^{-17}$ m$^3$, which is $\sim$10$^{-7}$ of particle volume. Thus, although our model is different from their experiments (e.g., density), the results give the comparable increase of the repose angle. Compared with the Earth, due to the small gravity, the angle of repose increases by a few degrees in the case of Mars ($g=3.7$ ms$^{-2}$). When the volume of liquid bridge is 5$\times$10$^{-19}$-2$\times$10$^{-18}$ m$^{3}$, the repose angle can reach the slope around the starting area of RSL ($\sim$35-40 degrees) (McEwen et al., 2011; Schmidt et al., 2017; Schaefer et al., 2019; Tebolt et al., 2019). 

\begin{figure}[t]
\centering
\includegraphics[width=12cm] {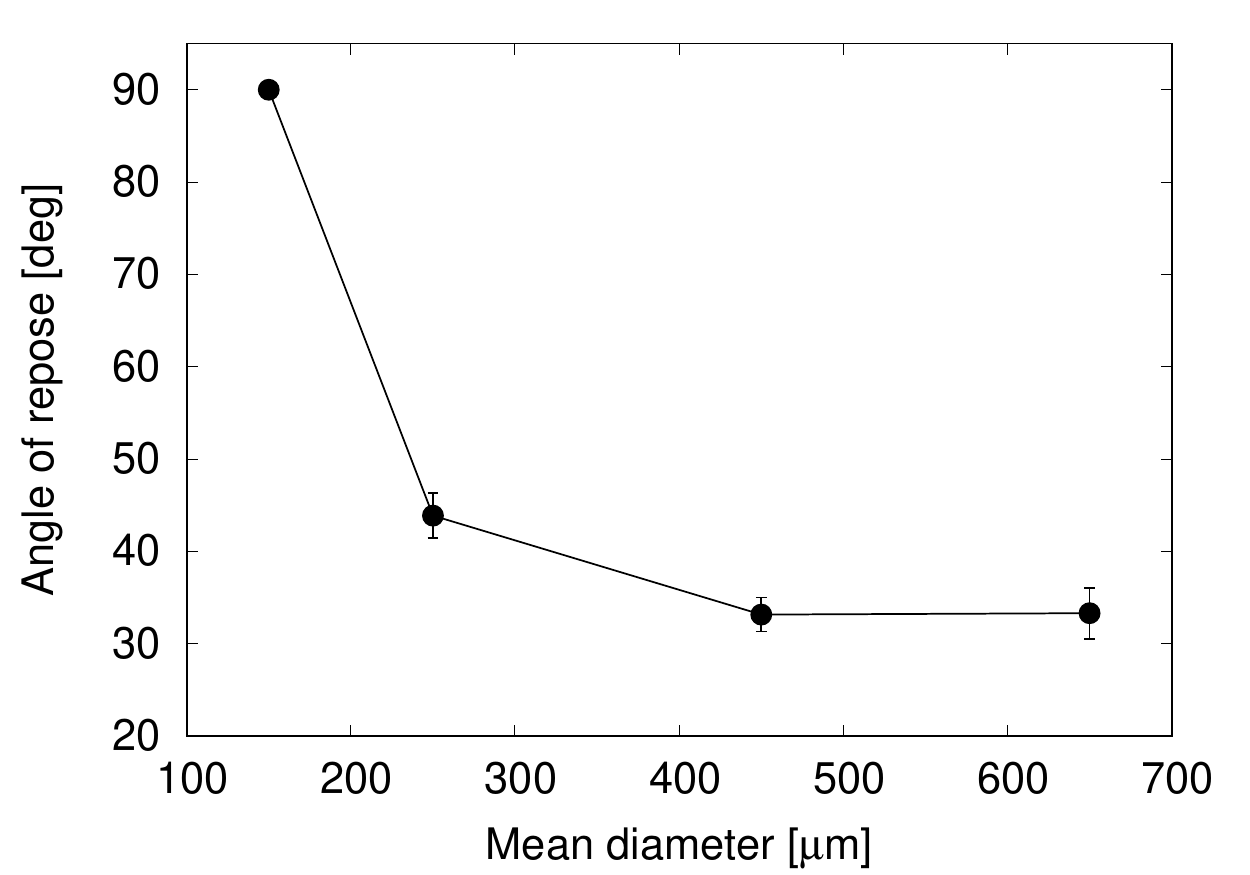}
\caption{Angle of repose at $V$=2$\times$10$^{-18}$ m$^3$ and $g$=3.7 ms$^{-2}$ as a function of mean diameter of particles. Error bars are 1$\sigma$ of standard deviation of 10 calculations.}
\label{fig3}
\end{figure}

Fig. \ref{fig3} shows the angle of repose as a function of mean particle diameter at $V=2\times10^{-18}$ m$^{3}$ and $g=3.7$ ms$^{-2}$. While the angle of repose little depends on particle diameter for dry grains, in the cases of wet particles, the angle is strongly affected by the particle size. Cohesion force works to resist the gravitational force, which increases with $R^3$. Thus, at constant $V$, gravity effect increases as the diameter increases, which results in the repose angle of dry grains ($\sim$30 degrees). Particles did not flow at 150 $\mu$m of mean diameter. In this case, although the definition of angle of repose is not valid, we assumed that $\theta_r$ is 90 degrees. In our calculations, if $V\sim$2$\times$10$^{-17}$ m$^3$, $\theta_r$ becomes $\sim$38 degrees even at 600-700 $\mu$m. Thus, if the ratio between a volumes of liquid bridge and a particle is the order of 10$^{-7}$, the angle of repose increases to the slope of the RSL starting area. 

\section{Conclusion}
It has been suggested that the seasonal cycle of temperature affects the formation of RSL (e.g., Stillman et al., 2014, 2016, 2017). Here, from the seasonal humidity change on Mars (Davila et al. 2010; P\'al et al., 2019) and terrestrial analog ("Saren" in Tottori Sand Dune) (Bitou and Kodama, 2011), we suggested that the recurrence of RSL may be caused by the seasonal humidity cycle as well as the temperature. If small amount of liquid is generated between particles by deliquescence or capillary condensation at high humidity season, as temperature increases and humidity decreases, wet particles on RSL sites begin to dry, which results in flow of sand or dust to downslope. This sand flow gradually increments depending on the magnitude of evaporation (drying), and finally the avalanche stops at the repose angle of dry particles ($\sim$30 degrees). This process may be the formation of RSL. When humidity increases and temperature decreases, particles are wet and can pile on RSL, which erases RSL and recharges sand for the next season. 

From 2D DEM simulations to evaluate the repose angle and the volume of liquid bridge, if the ratio of liquid bridge volume to grain volume is the order of 10$^{-7}$, the angle of repose reached the slope at starting area of RSL (35-40 degrees). This amount of liquid volume is much less than the required volume to generate RSL by liquid water itself. Thus water budget such as aquifer is not required. In addition, if the deposition of wet particles erases RSL, this mechanism can explain the seasonal cycle of RSL, which is a large problem for dry mechanism. 

Because evaporation rate of liquid depends on local temperature and humidity change, the seasonal and the diurnal climate of each RSL site are important. For example, around Argyre basin, increment season of several RSL sites is longer compared with other area (Stillman and Grimm, 2018; Vincendon et al., 2019). This may be because humidity in this area is lower as implied from the humidity maps by P\'al et al. (2019), which can induce granular flow for a longer time. In equatorial region, the cycle of RSL changes with the direction of slope (Stillman et al., 2017). Meta-stability of liquid also affects the state of liquid (Yang et al., 2008; Nosonovsky and Bhushan, 2008). Thus, further studies are required for the localized Martian climate change and the meta-stability of water. If deliquescence of salts causes the liquid bridge between particles, component of salts is also an important factor. As a further work, we are planning to perform experiments to evaluate the relationship between the angle of repose and humidity using Martian simulant.

\section*{Acknowledgements}
This work was supported by a JSPS Research Fellowship.

\newpage
\begin{table}[htbp]
\caption{Physical parameters and values. $m_{min}$ is the minimum mass of particle.}
\label{tab1}
\begin{tabular}{lcccc} \hline
Parameter&Symbol&Value&Unit&Reference\\ \hline
Number of particles&&800&\\
Density of particles&$$&2000&kg m$^{-3}$&Chojnacki et al. (2016)\\
Normal degrees constant&$k$&10$^6$&N m$^{-1}$&Nakashima et al. (2011)\\
Tangential degrees constant&$k_t$&$0.25k$&N m$^{-1}$&Nakashima et al. (2011)\\
Normal damping coefficient&$\eta$&1.8$\sqrt{m_{min}k}$&kg s$^{-1}$\\
Tangential damping coefficient&$\eta_t$&1.8$\sqrt{m_{min}k_t}$&kg s$^{-1}$\\ 
Surface tension&$\gamma$&0.073&N m$^{-1}$&Liu et al. (2011)\\
Contact angle&$\phi$&20&deg&Roy et al. (2016)\\
\hline
\end{tabular}
\end{table}

\newpage

\newpage

\newpage

\newpage

\newpage

\end{document}